
\noindent \hfil {\bf {SCALE-INVARIANT PHASE SPACE AND THE CONFORMAL GROUP}}
\bigskip
\noindent \hfil JAMES T. WHEELER

\noindent \hfil {\it {Department of Physics, Utah State University, Logan, UT
84322}}
\medskip
\noindent 1.  Introduction
\smallskip
Certain difficulties associated with any theory of quantum gravity stem from
the different mathematical framing of our gravity and quantum theories.
While general relativity is based on a real, deterministic, differentiable
manifold structure, the elements of quantum theory are complex-valued and
probabilistic.  Consider two ways to reconcile these differences by placing
one of the theories into the context of the other:
\smallskip
\item{A.}  Find an abstract quantum system (such as quantum string) with
general coordinate invariance and a massless, spin-2 state.  Gravitation may
emerge in terms of  excited states of the system.
\item{B.}  Find a curved geometry (necessarily non-Riemannian) which
naturally places the same restrictions on measurement as does quantum theory.
Quantum theory then emerges from the geometry.
\smallskip
In this work and elsewhere [1, 2] we display some progress toward approach B.
The geometric arena for our approach is the biconformal fiber bundle, defined
as the gauge bundle of the 4-dim conformal group over an 8-dim base space.
Here, we show how biconformal space can be used to characterize the dynamics
of Hamiltonian systems.  Specifically, we show that the classical Hamiltonian
dynamics of a point particle is equivalent to the specification of a 7-dim
hypersurface in flat biconformal space and the consequent necessary existence
of a set of preferred curves.  The central importance of this result is in
definitively establishing the physical interpretation of conformal gauging.
Given the result of section 3, that a classical Hamiltonian system generates
a class of biconformal spacetimes, there can remain little doubt as to the
meaning of the geometric variables.

In [1, 2], we go on to derive the general solution for flat biconformal space
and show how that solution predicts the electromagnetic potential including
its form, $\bf {\alpha }(x)$ as a vector field
 on 4-dim spacetime, its gauge dependence
$\bf {\alpha}'(x) = \bf {\alpha}(x) + \bf {d}\phi$, its minimal coupling
$p_{a} \longrightarrow p_{a} - \lambda \alpha_{a}$, and the correct equation
of motion for a charged particle moving under its influence.  This second
work therefore provides part of the consistent realization of Weyl's goal of
expressing the electrodynamics of a charged particle in terms of a Weyl
geometry [3].  Full realization of the goal requires in addition the
specification of Maxwell's field equations in terms of the biconformal
variables.  Reference [2] contains an examination of measurement when the
biconformal space is no longer dilationally flat.  We show there that the
rules of quantum mechanics provide a consistent way to make measurements of
physical size change in dilationally curved biconformal space.
\smallskip
To provide the background setting for the present discussion of Hamiltonian
dynamics in this new geometric context,  we describe below
the conformal group and its gauging to give the structure
equations of biconformal space, and display the flat solution for a local
frame field satisfying those structure equations.  Then, we establish our
central claim in three steps.  First, we show that the classical Hamiltonian
description of a point particle defines a class of 7-dim differential
geometries with structure equations of manifestly biconformal type.  Next, we
show that the introduction of an independent eighth coordinate in place of
the Hamiltonian allows, up to locally symplectic changes of basis, definition
of a unique flat biconformal geometry.  Finally, specializing to the
hypersurface determined by setting the eighth coordinate equal to any given
Hamiltonian, we show the necessary existence of a preferred set of curves in
the biconformal space satisfying the Hamiltonian equations of motion.
This unique equivalence between the hypersurfaces in
 flat biconformal geometry and the Hamiltonian
 system provides a clear physical interpretation of the geometric variables
of biconformal space.
\medskip
\noindent 2.  Biconformal space
\smallskip
Recall the 15 transformations comprising the 4-dim conformal group:
\smallskip
\hskip 0.5in 1.  Lorentz transformations (6)   \par
\hskip 0.5in 2.  Translations (4)  \par
\hskip 0.5in 3.  Inverse translations (4)   \par
\hskip 0.5in 4.  Dilation (1)      \par
\smallskip
\noindent These preserve ratios of infinitesimal lengths and therefore permit
global rescalings, which we identify as changes of units -- an obvious
symmetry of physical laws.  It is important to clearly distinguish global or
local changes of units, i.e., Weyl's original gauge transformations, from the
{\it {physical}} size changes that are the consequence of the local
dilational gauge vector, the Weyl vector, having nonvanishing curl.  The
clear imperative for any physical theory based on the gauging of the
conformal group is that any such physical size changes be given an
interpretation consistent with known measurements.  These issues will be
addressed further below, and in more detail in [2].

Extending the global conformal transformations to local symmetries gives
conformal geometry.  The extension is accomplished by introducing one gauge
1-form, $\omega_{B}^{A}$, for each generator of the group.  We choose the
linear $O(4,2)$ representation for our notation, with $(A, B, \dots) = (0, 1,
\ldots, 5)$.  Letting boldface or Greek symbols denote forms and $(a, b,
\dots) = (1, \ldots, 4)$, $\omega_{B}^{A}$ has four independent Lorentz
invariant parts:  the {\it {spin connection}}, $\omega_{b}^{a}$, the {\it
{solder form}}, $\omega_{0}^{a}$, the {\it {co-solder form}},
$\omega_{a}^{0}$, and the {\it {Weyl vector}}, $\omega_{0}^{0}$.
Orthonormality of the basis requires
$$ \eqalignno{
\omega_{b}^{a} &= - \eta_{bc} \eta^{ad} \omega_{d}^{c} \cr
\omega_{0}^{5} &= \omega_{5}^{0} = 0   \cr
\omega_{5}^{a} &= - \eta^{ab} \omega_{b}^{0} \cr
\omega_{a}^{5} &= - \eta_{ab} \omega_{0}^{b} & (1)\cr}$$

\noindent The structure constants of the conformal Lie algebra now lead
immediately to the Maurer-Cartan structure equations.  These are simply
$${\bf {d\omega}}_{B}^{A} = - {\bf {\omega}}_{B}^{C} \wedge {\bf
{\omega}}_{C}^{A} + {\bf {\Omega}}_{B}^{A} \eqno (2) $$
When broken into parts based on Lorentz transformation properties, eq.(2)
gives:
$$\eqalignno{
{\bf {d\omega}}_{b}^{a} &= - \omega_{b}^{c} \wedge \omega_{c}^{a} -  {\bf
{\omega}}_{b}^{0} \wedge {\bf {\omega}}_{0}^{a} - \eta_{bc}\eta^{ad} {\bf
{\omega}}_{0}^{c} \wedge {\bf {\omega}}_{d}^{0} + {\bf {\Omega}}_{b}^{a} \cr
{\bf {d\omega}}_{0}^{a} &= - {\bf {\omega}}_{0}^{0} \wedge {\bf
{\omega}}_{0}^{a} -  {\bf {\omega}}_{0}^{b} \wedge {\bf {\omega}}_{b}^{a} +
{\bf {\Omega}}_{0}^{a} \cr
{\bf {d\omega}}_{a}^{0} &= - {\bf {\omega}}_{a}^{0} \wedge {\bf
{\omega}}_{0}^{0} -  {\bf {\omega}}_{a}^{b} \wedge {\bf {\omega}}_{b}^{0} +
{\bf {\Omega}}_{a}^{0} \cr
{\bf {d\omega}}_{0}^{0} &=  -  {\bf {\omega}}_{0}^{a} \wedge {\bf
{\omega}}_{a}^{0} + {\bf {\Omega}}_{0}^{0}  &  (3) \cr} $$
Notice that if we set $\omega_{a}^{0} = \omega_{0}^{0} = 0$, we recover the
usual structure equations of 4-dim Riemannian geometry.  The meaning of
eqs.(2) depends on the choice of the base manifold.  For a 4-dim base space
the curvatures would be required to take the form
$${\bf {\Omega}}_{B}^{A} =  {\bf {\Omega}}_{Bcd}^{A} \> {\bf
{\omega}}_{0}^{c} \wedge {\bf {\omega}}_{0}^{d}  \hskip 0.5in  (A, B = 0, 1,
2, 3, 4) \eqno (4) $$
The fibres would then span the inhomogeneous Weyl group consisting of Lorentz
transformations, (co-\nobreak) translations, and dilations.  This gauge
theory has been extensively researched [4].  Quite generally, the four
1-forms $\omega_{a}^{0}$ are auxiliary and may be algebraically removed from
the problem [5].  The resulting geometry is always a Weyl geometry
(Riemannian plus dilations) and the action reduces to a linear combination of
the square of the conformal curvature and the square of the field strength of
the Weyl vector.  Thus, no additional structure is gained in the 4-dim
formulation by extending from the 11-parameter Weyl group to the 15-parameter
conformal group.

However, the approach to be followed here differs strikingly from this
standard picture.  The conformal group, $\cal {C}$, posesses eight
translational generators, distinguishable from the remaining generators by
their fixed points.  These remaining generators form the isotropy subgroup,
${\cal {C}}_{0}$, and the natural base space ${\cal {C}}/ {\cal {C}}_{0}$ is
8-dimensional.  The curvatures therefore take the form
$${\bf {\Omega}}_{B}^{A} =  {\bf {\Omega}}_{Bcd}^{A} \> {\bf
{\omega}}_{0}^{c} \wedge {\bf {\omega}}_{0}^{d} + {\bf {\Omega}}_{Bd}^{Ac} \>
{\bf {\omega}}_{c}^{0} \wedge {\bf {\omega}}_{0}^{d}   +  {\bf
{\Omega}}_{B}^{Acd}\> {\bf {\omega}}_{c}^{0} \wedge {\bf {\omega}}_{d}^{0}
\eqno (5) $$
with Lorentz transformations and dilations as the fibre symmetry.  The 8-dim
base manifold spanned by $\omega_{0}^{a}$ and $\omega_{a}^{0}$ will be called
{\it {biconformal space}} and the full fiber bundle the biconformal bundle.

Notice that the 8-dim form of the curvature allows some of the terms in the
structure equations to be incorporated into the curvature.  Specifically, we
define
$$\eqalignno{
{\overline{\bf {\Omega}}}_{0}^{0} &\equiv  -  {\bf {\omega}}_{0}^{a} \wedge
{\bf {\omega}}_{a}^{0} + {\bf {\Omega}}_{0}^{0} \cr
{\overline{\bf {\Omega}}}_{b}^{a} &\equiv  -  {\bf {\omega}}_{b}^{0} \wedge
{\bf {\omega}}_{0}^{a} -  \eta_{bc} \eta^{ad} {\bf {\omega}}_{0}^{c} \wedge
{\bf {\omega}}_{d}^{0} + {\bf {\Omega}}_{b}^{a}  & (6) \cr}$$
so that the corresponding structure equations reduce to
$$\eqalignno{
{\overline{\bf {\Omega}}}_{0}^{0} &= {\bf {d\omega}}_{0}^{0} \cr
{\overline{\bf {\Omega}}}_{b}^{a} &= {\bf {d\omega}}_{b}^{a} + {\bf
{\omega}}_{b}^{c} \wedge {\bf {\omega}}_{c}^{a} & (7) \cr}$$
If the curvatures of the solder and co-solder forms (i.e., the torsion and
co-torsion) leave the remaining two structure equations in involution then
the full set of structure equations are those of a pair of scale- and
Lorentz-conjugate 4-dim Weyl geometries.  The scale and Lorentz
transformations have inverse effects on the two 4-dim subspaces, with
covariant, weight $+1$ vectors in one space corresponding to contravariant,
weight $-1$ vectors in the other.  These transformations are not independent,
e.g., if a given Lorentz transformation acts on one of the 4-dim subspaces,
the inverse of the {\it {same}} Lorentz transformation acts on the conjugate
space.

It can be shown [1] that for vanishing curvature, ${\bf {\Omega}}_{B}^{A} =
0$, there locally exist coordinates $x^{a}, y_{a}$ such that
$$\eqalignno{
{\bf {\omega}}_{0}^{0} &= \alpha_{a}(x) {\bf {d}}x^{a} - y_{a} {\bf {d}}
x^{a} \equiv W_{a} {\bf {d}} x^{a} \cr
{\bf {\omega}}_{0}^{a} &=  {\bf {d}} x^{a} \cr
{\bf {\omega}}_{a}^{0} &=  {\bf {d}} y_{a} - (\alpha_{a,b} + W_{a}W_{b} -
{\scriptstyle {1 \over 2}} W^{2} \eta_{ab}) {\bf {d}} x^{b} \cr
{\bf {\omega}}_{b}^{a} &=  ( \delta_{d}^{a} \delta_{b}^{c} - \eta^{ac}
\eta_{bd}) W_{c} {\bf {d}} x^{d} &(8) \cr} $$
where  $\alpha_{a,b}$ denotes the partial of $\alpha_{a}$ with respect to
$x^{b}$.  There are several points to note about these solutions:
\item{(1)}  The Weyl vector, $W_{a}$, while having only four nonvanishing
components in this basis, is actually a function of $y_{a}$ as well.  This
necessary dependence is the reason for the failure of Weyl's original
scale-invariant theory of electromagnetism [2].
\item{(2)}  The solutions hold for vanishing conformal curvatures, not for
the barred curvatures.  Vanishing barred curvatures would lead to a flat pair
of conjugate Riemannian spacetimes, while vanishing conformal curvatures
correspond to constant curvature Weyl spacetimes.
\item{(3)} This form of the solution is preserved by 4-dim gauge
transformations, $\phi(x)$.  The gauge transformation  must be associated
with the undetermined vector field $\alpha_{a} (x)$.   For later use we note
that ${\bf {\omega}}_{0}^{a} \wedge {\bf {\omega}}_{a}^{0} = {\bf {d}} x^{a}
\wedge {\bf {d}} y_{a} - \alpha_{a,b} {\bf {d}} x^{a} \wedge {\bf {d}} x^{b}$
reduces to ${\bf {\omega}}_{0}^{a} \wedge {\bf {\omega}}_{a}^{0} = {\bf {d}}
x^{a} \wedge {\bf {d}} y_{a}$ if and only if $\alpha_{a}$ is pure gauge.
\medskip
\noindent 3.  A geometry for Hamiltonian mechanics.
\smallskip
Leaving biconformal geometry for the moment, we turn to a geometric approach
to classical Hamiltonian dynamics.  We first show that the action of a
classical system may be used to generate biconformal spaces.  Then, we give a
unique prescription for generating a flat biconformal space.   Begin with the
Hilbert form
$${\bf {\omega}} =  H {\bf {d}} t - p_{i} {\bf {d}} x^{i} \eqno (9)$$
where $H = H(p_{i},x^{i},t)$.  The integral of $\bf {\omega}$ is the action
functional.  The exterior derivative of $\bf {\omega}$ may always be
factored:
$$\eqalignno{
{\bf {d \omega}} &= -{\partial H \over \partial x^{i}} {\bf {d}} x^{i} \wedge
{\bf {d}} t - {\partial H \over \partial p_{i}} {\bf {d}} p_{i} \wedge {\bf
{d}} t  - {\bf {d}} x^{i} \wedge {\bf {d}} p_{i} \cr
&= - ( {\bf {d}} x^{i}  -  {\partial H \over \partial p_{i}} {\bf {d}} t )
\wedge ( {\bf {d}} p_{i}  +  {\partial H \over \partial x^{i}}  {\bf {d}} t )
&(10) \cr} $$
Therefore, if we define
$$\eqalignno{
{\bf {\omega}}^{i} &\equiv ({\bf {d}} x^{i} - {\partial H \over \partial
p_{i}} {\bf {d}} t) \cr
{\bf {\omega}}_{i} &\equiv ({\bf {d}}p_{i} + {\partial H \over \partial
x^{i}} {\bf {d}} t)  &(11) \cr }$$
then we can write
$${\bf {d\omega}} =  -  {\bf {\omega}}^{i} \wedge {\bf {\omega}}_{i}  \eqno
(12) $$
This factoring is clearly preserved by local symplectic transformations of
the 6-basis $(\omega^{i}, \omega_{i})$, as well as reparameterizations of the
time.  Obviously these transformations include the usual canonical
transformations of coordinates as a special case.  One class of such allowed
transformations of basis is achieved by the addition of $c_{ij}{\bf {d}}
x^{j}$, with $c_{ij} = c_{ji}$, to ${\bf {\omega}}_{i}$  Here we take $c_{ij}
= 0$, but in section 4 we show the existence of a unique choice of the 4-dim
extension of $c_{ij}$ that leads to flat biconformal space.

Continuing, we wish to show that any choice for ${\bf {\omega}}^{i}$ and
${\bf {\omega}}_{i}$ leads to a {\it {dilationally}} flat biconformal space.
Taking the exterior derivatives of ${\bf {\omega}}^{i}$ and ${\bf
{\omega}}_{i}$ we find:
$$\eqalignno{
{\bf {d\omega}}^{i} &=  {\partial^{2}H \over \partial x^{j} \partial p_{i}}
{\bf {d}} x^{j} \wedge {\bf {d}} t   +   {\partial^{2}H \over \partial p_{j}
\partial p_{i}}  {\bf {d}} p_{j} \wedge  {\bf {d}} t  \cr
{\bf {d\omega}}_{i} &=  - {\partial^{2}H \over \partial x^{i} \partial p_{j}}
{\bf {d}} p_{j} \wedge {\bf {d}} t   -   {\partial^{2}H \over \partial x^{j}
\partial x^{i}}  {\bf {d}} x^{j} \wedge  {\bf {d}} t .  & (13) \cr}$$
If we define
$$\eqalignno{
{\bf {\omega}}_{j}^{i} &\equiv -{\partial^{2}H \over \partial x^{j} \partial
p_{i}} {\bf {d}} t  \cr
{\bf {\Omega}}^{i}  &\equiv  {\partial^{2}H \over \partial p_{j} \partial
p_{i}} {\bf {d}} p_{j} \wedge  {\bf {d}} t  \cr
{\bf {\Omega}}_{i}  &\equiv  -{\partial^{2}H \over \partial x^{j} \partial
x^{i}}  {\bf {d}} x^{j} \wedge {\bf {d}} t &(14) \cr } $$
then eqs.(13) are simply
$$\eqalignno{
{\bf {d\omega}}^{i} &= - {\bf {\omega}}^{j} \wedge {\bf {\omega}}_{j}^{i}  +
{\bf {\Omega}}^{i} \cr
{\bf {d\omega}}_{i} &= - {\bf {\omega}}_{i}^{j} \wedge {\bf {\omega}}_{j}  +
{\bf {\Omega}}_{i} &(15)  \cr } $$
Finally differentiating ${\bf {d \omega}}_{j}^{i}$, using the fact that ${\bf
{d \omega}}_{j}^{k} \wedge {\bf {d \omega}}_{k}^{i} = 0$ and defining the
final curvature as
$${\bf {\Omega}}_{j}^{i} \equiv {\bf {d\omega}}_{j}^{i} +  (\delta_{k}^{i}
\delta_{j}^{l} -  \delta_{jk} \delta^{il} )  {\bf {\omega}}^{k} \wedge {\bf
{\omega}}_{l} \eqno (16)$$
gives the complete set of structure equations: \smallskip
$$\eqalignno{
{\bf {d\omega}}_{j}^{i} &= - {\bf {\omega}}_{j}^{k} \wedge {\bf
{\omega}}_{k}^{i} - {\bf {\omega}}^{i} \wedge {\bf {\omega}}_{j} +
\delta_{jk} \delta^{il} {\bf {\omega}}^{k} \wedge {\bf {\omega}}_{l} + {\bf
{\Omega}}_{j}^{i} \cr
{\bf {d\omega}}^{i} &= - {\bf {\omega}}^{j} \wedge {\bf {\omega}}_{j}^{i} +
{\bf {\Omega}}^{i} \cr
{\bf {d\omega}}_{i} &= -  {\bf {\omega}}_{i}^{j} \wedge {\bf {\omega}}_{j} +
{\bf {\Omega}}_{i} \cr
{\bf {d\omega}}\> &=  -  {\bf {\omega}}^{i} \wedge {\bf {\omega}}_{i} &(17)
\cr}$$
The final equation, for ${\bf {d\omega}}$, and the dependence of the
curvatures on {\it {both}} ${\bf {\omega}}^{i}$ and ${\bf {\omega}}_{i}$
clearly show this to be a subspace of a biconformal space.  Since we may
expand
$$\omega = H{\bf {d}}t - p_{i} {\bf {d}}x^{i} =  (H - p_{i}{\partial H \over
\partial p_{i}}) {\bf {d}}t - p_{i} \omega^{i} \eqno (18)  $$
we could also have included the Weyl vector terms present in eqs.(3) in the
definitions of the torsion and co-torsion.  Therefore, since $\omega_{i}$
depends on ${\bf {d}}p_{i}$, biconformal space may be viewed as a form of
scale-invariant phase space.  We note that this interpretation agrees in some
ways with earlier proposals [6] relating phase space and Weyl geometry.
These proposals, however, lack the full geometric structure of conformal
gauge theory, do not demonstrate the intrinsically biconformal structure of
Hamiltonian systems, and use a different inner product than that proposed in
[2].

We next show the relationship of the geometry above to classical mechanics.

In the 7-dim geometry defined above, the combined involution of ${\bf
{\omega}}^{i}$ and ${\bf {\omega}}_{i}$ in eqs.(14, 15) allows us to set
${\bf {\omega}}^{i} = {\bf {\omega}}_{i}  =  0$, thereby singling out a
fibration of the bundle by 1-dim subspaces, i.e., the classical paths of
motion.  These simply give Hamilton' s equations of motion:
$$ \eqalignno{  {\bf {d}} x^{i}  &=  {\partial H \over \partial p_{i}} {\bf
{d}} t \cr
{\bf{d}} p_{i}  &=  -  {\partial H \over \partial x^{i}} {\bf {d}} t  & (19)
\cr } $$
The Frobenius theorem guarantees the existence of solutions to these
equations for the paths.  The structure equations then reduce to
$$ {\bf {d\omega}}_{j}^{i} = {\bf {d\omega}} = 0 \eqno (20) $$
which are identically satisfied on curves.

On the full bundle, the condition that ${\bf {\omega}}$ be exact is the
Hamiltonian-Jacobi equation, since we may write ${\bf {\omega}} = {\bf {d}}
S$.  Substituting for ${\bf {\omega}}$ and expanding ${\bf {d}} S$ then gives
$$-H{\bf {d}}t + p_{i} {\bf {d}} x^{i} = {\partial S \over \partial x^{i} }
{\bf {d}} x^{i} + {\partial S  \over  \partial p_{i} } {\bf {d}} p_{i} +
{\partial S \over \partial t} {\bf {d}} t \eqno (21)$$
so that
$${\partial S \over \partial p_{i} }=0 \eqno (22)$$
$${\partial S \over \partial x^{i} }= p_{i} \eqno (23)$$
and
$$H({\partial S \over \partial p_{i} },x^{i},t) = -{\partial S \over \partial
t}. \eqno (24)$$
Therefore, since ${\bf {\omega}}$ is the Weyl vector of the generated
biconformal space, the Hamilton-Jacobi equation holds if and only if the Weyl
vector is pure gauge, ${\bf {\omega}} = {\bf {d}}S$.  A gauge transformation
reduces the Weyl vector to zero.  Since the dilational curvature is also
always zero for the geometry built from a Hamiltonian system, we also have
$$\omega^{i} \wedge \omega_{i} = 0, \eqno (25)$$
implying three linear dependences between these six solder forms.  Together
with the linear dependence of ${\bf {d}}H$, the Hamilton-Jacobi equation
therefore specifies a 4-dim subspace of the full biconformal space.
\medskip
\noindent 4.  Flat biconformal space and the Hamiltonian geometry
\smallskip
Next, we show how to specify a unique {\it {flat}} biconformal space for a
given Hamiltonian system.  This is achieved by first placing the Hamiltonian
geometry in an 8-dim setting, then by being careful in the choice of the
solder and co-solder forms.

First, replace  $H =  H(p_{i},x^{i},t)$ by an independent variable, $-p_{4}$.
Then
$${\bf {\omega}_{0}^{0}} =  -p_{a} {\bf {d}} x^{a} \eqno (26)$$
and
$${\bf {d \omega}_{0}^{0}} = -  {\bf {d}} x^{a} \wedge {\bf {d}} p_{a}, \eqno
(27)$$ where we have reserved the symbols $\omega, \omega_{i}$ and
$\omega^{i}$ for the special case when $p_{4} = - H(p_{i},x^{i},t)$.
This time, we make our identification of the solder and co-solder forms
$\omega_{0}^{a}$ and $\omega_{a}^{0}$ by comparing the expression  for ${\bf
{d}} \omega_{0}^{0}$ to the flat biconformal solution.  There is  clearly a
{\it {unique}} extension to a flat biconformal space, given by setting
$\alpha_{[a,b]} = 0$ in the general flat solution, as noted above.  A scale
transformation then removes $\alpha_{a}$ altogether so that, in terms of the
usual phase-space coordinates, the frame field of the flat biconformal
extension becomes
$$\eqalignno{
{\bf {\omega}}_{0}^{0} &=  - p_{a} {\bf {d}} x^{a} \cr
{\bf {\omega}}_{0}^{a} &=  {\bf {d}} x^{a} \cr
{\bf {\omega}}_{a}^{0} &=  {\bf {d}} p_{a} - (p_{a}p_{b} - {\scriptstyle {1
\over 2}} p^{2} \eta_{ab}) {\bf {d}} x^{b}  \cr
{\bf {\omega}}_{b}^{a} &= (\delta_{d}^{a} \delta_{b}^{c} - \eta^{ac}
\eta_{bd}) p_{c} {\bf {d}} x^{d} & (28) \cr  }$$

If we now restrict to the hypersurface $ p_{4} = - H(p_{i},x^{i},t)$ we
recover the Hamiltonian system, and have shown it to lie in a unique flat
biconformal space.  Note that the symmetric coefficient
$$c_{ab} = p_{a}p_{b} - {\scriptstyle {1 \over 2}} p^{2} \eta_{ab} \eqno (29)
$$
provides a symplectic change of basis with respect to the manifestly closed
and nondegenerate 2-form, ${\bf {d}} \omega_{0}^{0} = - \omega_{0}^{a} \wedge
\omega_{a}^{0} = -{\bf {d}}x^{a} \wedge {\bf {d}}p_{a}$.  The differential of
$\omega_{0}^{0}$ is still seen to factor into the form of eq.(10)  either
directly by differentiation or by substitution of $ p_{4} = -
H(p_{i},x^{i},t)$ into $- {\bf {\omega}}_{0}^{a} \wedge {\bf
{\omega}}_{a}^{0}$.  However, while the involution of eq.(13) for
$\omega_{i}$ and $\omega^{i}$ still holds it does not mean that
$\omega_{0}^{a}$ and $\omega_{a}^{0}$ vanish.  Instead, the classical curves
are found by first writing the frame field in terms of $\omega^{i}$ and
$\omega_{i}$:
$$\eqalignno{
\omega_{0}^{0} &= (H - p_{i}{\partial H \over \partial p_{i}}) {\bf {d}}t -
p_{i} \omega^{i} \cr
\omega_{0}^{i} &= \omega^{i} + {\partial H \over \partial p_{i}} {\bf {d}}t
\cr
\omega_{0}^{4} &= {\bf {d}}t \cr
\omega_{i}^{0} &= \omega_{i} - (p_{i}p_{j} + {\scriptstyle {1 \over 2}}
(H^{2} - p^{2}) \delta_{ij}) \omega^{j} + (p_{i}H - {\partial H \over
\partial x^{i}} - {\scriptstyle {1 \over 2}} (H^{2} - p^{2})
\delta_{ij}{\partial H \over \partial p_{j}}) {\bf {d}} t \cr
\omega_{4}^{0} &= (Hp_{i} - {\partial H \over \partial x^{i}}) \omega^{i} +
(H{\partial H \over \partial p_{i}} p_{i} -  {\scriptstyle {1 \over 2}}
(H^{2} - p^{2}) - {\partial H \over \partial x^{i}}{\partial H \over \partial
p_{i}})  {\bf {d}} t  \cr
\omega_{j}^{i} &= (\delta_{n}^{i} \delta_{j}^{m} - \delta^{im} \eta_{jn})
p_{m} (\omega^{n} + {\partial H \over \partial p_{n}}{\bf {d}}t)  \cr
\omega_{4}^{i} &= -H \omega^{i}+ (\delta^{ij}p_{j} - H{\partial H \over
\partial p_{i}}) {\bf {d}}t  \cr
\omega_{i}^{4} &= \delta_{ij} \omega_{4}^{j} & (30) \cr}$$
The simple example of a free particle is instructive.  Setting $\omega^{i} =
\omega_{i} = 0$ and $H^{2} = p^{2} + m^{2} \neq 0$, the equations above
reduce to:
$$\eqalign{
\omega_{0}^{0} &= {m^{2} \over H} \> {\bf {d}}t  \cr
\omega_{0}^{a} &= {\eta^{ab} p_{b} \over m^{2} } \> \omega_{0}^{0}  \cr
\omega_{a}^{0} &= {\scriptstyle {1 \over 2}}p_{a} \> \omega_{0}^{0}  \cr
\omega_{b}^{a} &= 0  \cr}$$
where use of  $\omega_{0}^{0}$ in place of ${\bf {d}}t$  simplifies the
expressions.  The solder and co-solder forms are proportional to the
displacement and momentum, respectively.

Also, we see that the involution required to specify the classical paths
necessarily exists.  Since the biconformal base space is spanned by the eight
forms ${\bf {d}}x^{a}$ and ${\bf {d}}p_{a}$, the only way the involution
could fail is if there was an independent  ${\bf {d}}t \wedge {\bf {d}}p_{4}$
term in the torsion or co-torsion of the $p_{4} = - H$ hypersurface.  But
since ${\bf {d}}H$ is given {\it {a priori}} in terms of the other seven
forms, this cannot happen.
\medskip
We conclude: {\it {the Hamiltonian dynamics of a point particle is equivalent
to the specification of a hypersurface, $y_{4} = y_{4}(y_{i},x^{a})$, in a
flat biconformal space, and the consequent necessary existence of preferred
curves in the hypersurface.}}
\medskip
The principal significance of this result is the unambiguous  identification
of the geometric quantities which arise in the gauging of the 15-dim
conformal group.  While previous treatments of this gauge theory always lead
to a 4-dim Weyl geometry in which the special conformal transformations
(i.e., the inverse translations) are auxiliary, the current approach includes
additional freedom beyond that of a Weyl geometry.  This extra freedom is now
seen to correspond to the inclusion of momentum variables in the physical
description.
\bigskip
The author thanks Y. H. Clifton and C. Torre for entertaining and useful
discussions.
\bigskip
\hfil References
\item{[1]} Wheeler, J. T., in preparation.
\item{[2]} Wheeler, J. T., in preparation.
\item{[3]} Weyl, H., Sitzung. d. Preuss. Akad. d. Wissensch. (1918) 465;
Weyl, H., {\it {Space-Time-Matter}}, Dover Publications, New York (1952).
\item{[4]} Ferber, A. and P.G.O. Freund, Nucl. Phys. {\bf {B122}} (1977) 170;
	Crispim-Romao, J., A. Ferber and P.G.O. Freund, Nucl. Phys. {\bf {B126}}
(1977) 429;
Kaku, M., P.K. Townsend and P. Van Nieuwenhuizen, Phys. Lett. {\bf {69B}}
(1977) 304.
\item{[5]} Wheeler, J. T., Phys Rev D{\bf {44}} (1991) 1769.
\item{[6]} Caianiello, E. R., M. Gasperini, E. Predazzi and G. Scarpetta,
Phys. Lett. {\bf {132A}}, 2, (1988) 82; Caianiello, E. R., A. Feoli, M.
Gasperini and G. Scarpetta, Intl. J.Theor. Phys., {\bf {29}}, 2, (1990), 131
and references therein.
\bye